\documentclass[11pt]{article}
\usepackage[dvips]{graphicx}
\usepackage{floatflt}
\usepackage{color}
\usepackage{colortbl}
\usepackage{cite}

\newcommand{\be}{\begin{equation}}
\newcommand{\ee}{\end{equation}}
\newcommand{\bean}{\begin{eqnarray}}
\newcommand{\eean}{\end{eqnarray}}
\newcommand{\bea}{\begin{eqnarray*}}
\newcommand{\eea}{\end{eqnarray*}}

\newcommand{\bc}{\begin{center}}
\newcommand{\ec}{\end{center}}

\suppressfloats[!]

\begin{document}


\pagestyle{empty}

\vspace*{3.0cm}

\begin{center}     
   
\LARGE{\bf\boldmath{Relativistic and Nonrelativistic Descriptions
       of Electron Energy Levels in a Static Magnetic Field}}

\end{center}

\vspace{1.0cm}
\begin{center}

H.J.~Schreiber$^1$ and N.B.~Skachkov$^2$\\

\end{center}

\bigskip \bigskip
\begin{center}
\small
\vspace{-0.5cm}
$^1$DESY, Platanenallee 6, D-15738 Zeuthen, Germany \\ [2mm]
$^3$JINR, Joliot Curie 6, 141980, Dubna, Moscow Region, Russia \\

\end{center}

\normalsize

\pagestyle{plain}
\pagenumbering{arabic}

\vspace{3.0cm}
\begin{center}
\section*{Abstract}
\end{center}
\noindent 
     The physical consequences of the relativistic and
     nonrelativistic  approaches to describe the energy
     levels of  electrons which propagate in a static
     homogeneous  magnetic field  are considered.
     It is shown that for a given strength of the magnetic
     field, the quantized energy levels of the electrons
     calculated by nonrelativistic and relativistic
     equations  differ substantially, up to few orders
     of magnitude for a magnetic field of about 1 Tesla.
     Experimental verification to resolve the 
     discrepancy would be very welcome.

\small
\vspace{30mm}
\noindent
E-mail: schreibe@ifh.de, skachkov@jinr.ru
\normalsize

\newpage


\section {Introduction}
\noindent
   The existence of quantized transverse energy levels of charged
  particles which propagate  in a static homogeneous magnetic
  field  was predicted by the nonrelativistic                          
  Schr\"odinger equation \cite{Land} and the relativistic
  Klein-Gordon  equation  for a scalar particle \cite{Page, Pleesset}. 
  Later, analogous expressions
  for transverse energy levels of electrons were
  found within the Dirac equation, see e.g. \cite{Akhi}.
  For more details we refer to 
   \cite{Sok-Tern, Sokolov-Ternov1, Sokolov-Ternov2, Bagr_etal}.
   Furthermore, exact solutions were also derived for
   somewhat  more complicated cases  when the electron pass through
   a combined static electric and
   magnetic field (Volkov type solutions, see also
   \cite{Redm, Berg}), and recently, this study was continued
   by finding the solution of the Dirac equation for a  
   superposition of a static homogeneous magnetic and 
   electric field including the anomalous magnetic moment
   of the electron \cite{R.A.Mel}.  It was also shown
   in ref.\cite{R.A.Mel} that by accounting for the electron
   anomalous magnetic moment the spin degeneracy of energy
   levels was removed.  So, according to all these results,
   an electron in a static magnetic field gyrates about, and
   moves along, the field lines and possesses quasi-atomic
   bound states with energy levels related to its gyrating
   motion in the plane normal to the velocity  vector.
    In the literature, these quasi-discrete resonance 
    states were discussed in connection with the motion of charged  particles
    in an external  magnetic field and accelerator physics,
    see \cite{Zioutas, Melik/Protv, BarMel, Milant} for example 
    \footnote{The text books  \cite{Sokolov-Ternov1, Sokolov-Ternov2, Bagr_etal}
              include a rather complete review of possible 
              applications starting from acceleration
              applications up to cosmology issues.}.
 
\vspace{3mm}
\noindent    
   The analytical solutions of the corresponding equations
   make clear what kind of
   physical effects may  appear during the transition
   from the nonrelativistic to the  relativistic formalism.
   The example of the Coulomb potential, which we mention
   in what follows, shows a nontrivial 
   physical sequence when passing from one case to the other.
      
\vspace{3mm}
\noindent    
   In this note we would like to draw the attention to the
   fact that the nonrelativistic  and relativistic approaches, 
   based on the Schr\"odinger, respectively, the Dirac equation,
   give different analytical expressions for transverse 
   energy levels of electrons in a static magnetic field.
   The solutions predict different dependencies of 
   these levels on the magnetic field, so that
   for a given field strength,
   rather different values for transverse
   energies are expected. 
   In order to check the validity of the different theoretical
   predictions it would be very welcome to measure
   the energy of photons emitted by electrons when transitions 
   from higher to lower orbits occur in an external magnetic field.
     
\vspace{3mm}
\noindent
     In Section 2 we discuss some  basic formulas which 
   define  the energy levels of an electron which
   traverses a static and uniform  magnetic field.
   Section 3 presents the numerical values 
   for transverse energy levels of electrons assuming a
   relatively strong magnetic field of 1 Tesla.
   Section 4 summarizes the discussion and proposes
   an experimental verification to decide which
   of the two theoretical concepts based on either
   the Dirac or Schr\"odinger equation is realized.
  
\vspace{3mm}
\noindent
Throughout this paper, the Gaussian system of units
    will be used.
     

\section{Energy levels and wave function of 
          electrons in a static magnetic field; solutions of the
          Schr\"odinger and Dirac equations}

  The energy spectrum obtained from the Schr\"odinger equation
  for electrons with spin $\frac{1}{2}$
  which are gyrating around the field lines of a static
  homogeneous  magnetic field , 
   ($\vec{H}=rot\vec{A}$)
  with, according to ref.\cite{Land}, the following choice of the
  4-vector of the electromagnetic potential  $A_{\mu}$:
   $A_{0}= A_{x}= A_{z}=0$,  $A_{y}=Hx$ 
   \footnote{Thus, $\vec{A}=Hx\vec{e_{y}}$, with 
              $\vec{e_{y}}$  the unit vector along 
      the y-axis,  and  $\vec{H}=(0,0,H_{z})$,
      with $H_{z}=H$, and the electron  
              momentum along the z-axis.},
 may be  written  as the sum of the longitudinal and transverse
  components \cite{LandLif}
\begin{equation}
    {E^{nonrel} = E_{z}^{nonrel} + E^{nonrel}_{T,\lambda}(n) ~, }
\end{equation} 
where  
\begin{equation}     
   {E_{z}^{nonrel}= \frac{p_{z}^{2}}{2m_{e}} }
\end{equation}
and 
\begin{equation}
{ E^{nonrel}_{T,\lambda}(n)= \hbar(\frac{eH}{2m_{e}c})(2n +1 +2\lambda)
 = (\mu_{B}^{e}H)(2n +1 +2\lambda) }~,
\end{equation}  
   the energy of the electron motion   
   in transverse direction
   \footnote{Here  $\hbar$  is the value
             of the  Plank constant h  divided  by $2\pi$, i.e.
             $\hbar=1.054 571 586(82) 10^{-34}$ J s.}.
   The transverse energy depends on the strength of the
   magnetic field $H$ and on the electron spin projection
   $\lambda$ onto the z-, i.e. the electron, direction.
   It possesses quantized  values labeled by
   the  main quantum number $n$ ($n=0,1,2,...$).
   In eq.(3), $m_{e}$ and $e$ denote the electron mass, 
   respectively, its charge, and
   $\frac{eH}{m_{e}c}=\omega_{c}$ is the cyclotron
   frequency. We  employed here  the definition of the
   Bohr magneton of an electron,
   $ \mu_{B}^{e} = \frac{e\hbar}{2m_{e}c}$.

\vspace{3mm}
\noindent
The energy levels,  defined by the relativistic Dirac 
  equation  for electrons in the same 
  magnetic field \cite{Akhi}, are  connected to the
  fourth  component of its 4-momentum vector,
   $P_{\mu}=(p^{0}, p^{x}, p^{y}, p^{z} )$,
   and are given as
\begin{equation}
  {c^2(p^{0})^2 \equiv  E^{2}_{\lambda}(n,p_{z} ) =
    E^{2}_{z} + E^{2}_{T,\lambda}(n)}~.
\end{equation}
    The first term 
\begin{equation}
   {E^{2}_{z}
    = m^{2}_{e}c^{4} +{p^{2}_{z}}c^{2}}
\end{equation}
   is the square of the relativistic energy of a free electron
      moving along the z-axis, whereas the second term 
      defines the  square of the relativistic  electron 
   transverse energy \footnote{The
            solution of the  Klein-Gordon equation
            \cite{Page, Pleesset} is
            obtained by omitting the term $ 2\lambda$ in eq.(3).}:
\begin{eqnarray}
E^{2}_{T,\lambda}(n) =
     (m_{e}c^{2})\hbar (\frac{eH}{m_{e}c})(2n +1 +2\lambda)
    =(m_{e}c^{2})\hbar\omega_{c}(2n +1 +2\lambda)\nonumber\\         
    = 2m_{e}c^{2}(\mu_{B}^{e}H)(2n +1 +2\lambda)~.
\end{eqnarray}    
 Comparing (3) and (6) the last equation may be expressed as
\begin{equation}
{E^{2}_{T,\lambda}(n) =
     2mc^{2}\cdot E^{nonrel}_{T,\lambda}(n)~.}
\end{equation}                               
Note that for the ground state
   (with $ n=0$) and  spin projection $\lambda=-\frac{1}{2}$, 
   the electron transverse energy is equal to zero in both 
   the nonrelativistic and relativistic approaches:
\begin{equation}
{E^{2}_{T,\lambda=-\frac{1}{2}}(n=0)\equiv E^{2}_{T,\lambda=-\frac{1}{2}}(0) =0 ~,}
\end{equation}
whereas the $n=0$ level with spin projection $\lambda=+\frac{1}{2}$
   has, according to (6), a non-zero transverse  energy squared of
\begin{equation}
   {E^{2}_{T,\lambda=+\frac{1}{2}}(n=0)
   = 2\hbar\omega_{c}(m_{e}c^{2})
   = 2~(2m_{e}c^{2})(\mu_{B}^{e}H)~. }
\end{equation}   

\vspace{3mm}
\noindent
So, we realize that \textit {for the ground state 
   with $\lambda=-\frac{1}{2}$, the relativistic expression for the
   total energy  of an electron in a static magnetic field}
   coincides  with \textit {the energy of a free electron}
\begin{equation}
  {E_{\lambda=-\frac{1}{2}}(n=0,p_{z})  = E_{z}}=
  \sqrt{ m^{2}_{e}c^{4} +{p^{2}_{z}}c^{2}}~,
\end{equation}                             
  which is however not the case for spin projection 
  $\lambda=+\frac{1}{2}$ .  One also notices from
  eqs.(6) and (9) that the state with the quantum 
  numbers $n=0$ and  $\lambda=+\frac{1}{2}$ has 
  the same transverse energy as the state with 
  $n=1$ and  $\lambda=-\frac{1}{2}$, i.e. 
\begin{equation}
  {E^{2}_{T,\lambda=-\frac{1}{2}}(n=1) 
  = E^{2}_{T,\lambda=+\frac{1}{2}}(n=0)
  =  2\hbar\omega_{c}(m_{e}c^{2})~
  = 2~(2m_{e}c^{2})(\mu_{B}^{e}H)~. }
\end{equation}   

\vspace{3mm}
\noindent
From eq.(6) one derives for the
   difference $\Delta{E^{2}_{T,\lambda}} (n+k|n)$
   of the square of two transverse energy levels $E^{2}_{T,\lambda}$,
   labeled as $n+k$ ($k=1, 2, ...$) and $n$,
   and identical spin projections $\lambda$, i.e.
   for the non-spinflip case, the following expression
\begin{eqnarray}
 {\Delta{E^{2}_{T,\lambda} (n+k|n)}\equiv 
    E^{2}_{T,\lambda}(n+k) - E^{2}_{T,\lambda}(n)  = 
     2k\hbar\omega_{c}(m_{e}c^{2})}   \nonumber\\
   = 4k(\mu_{B}^{e}H)(mc^{2})= 2ec\hbar kH  
    = k\Delta{E^{2}_{T,\lambda= -1/2} (1|0)~. }
\end{eqnarray}

\vspace{3mm} 
\noindent
The energy eigenvalues of eq.(4) in the 
   nonrelativistic limit might be
   expanded to \cite{Sokolov-Ternov1} 
\begin{eqnarray}
   E_{\lambda}(n,p_{z} ) =
   \sqrt{ E^{2}_{z} + E^{2}_{T,\lambda}(n) }=
   \sqrt{ m^{2}_{e}c^{4} +{p^{2}_{z}}c^{2} + 
   \hbar(\frac{eH}{m_{e}c})(m_{e}c^{2})(2n +1 +2\lambda)}   \nonumber\\
~~~~~~~~~~~~~~~~~ \approx {{m_{e}c^{2}} + ( \frac{p_{z}^{2}}{2m_{e}}) 
   + {(\mu_{B}H)}(2n +1 +2\lambda)= {m_{e}c^{2}} +  E^{nonrel}~. }
\end{eqnarray}
This equation clearly reveals the relationship between the relativistic energy
$E_{\lambda}(n,p_{z})$ to the nonrelativistic energy $E^{nonrel}$
or vice versa.

%

\section{Comparison of numerical values of transverse
          energy levels from the
          Schr\"odinger and Dirac equations}


   In the nonrelativistic Schr\"odinger case the transverse
    energy is, according to (3),
   proportional to the strength of the magnetic field H  
\begin{equation}
{E_{T,\lambda}^{nonrel}(n) \sim (\mu_{B}^{e}H)~, }
\end{equation}
   whereas in the relativistic case
   the transverse energy is, according to (6), proportional 
   to the  square root of H
\begin{equation}
{E_{T,\lambda}(n) = \sqrt { 2(m_{e}c^{2}) E^{nonrel}_{T,\lambda}(n)}
 \sim  \sqrt{ 2(m_{e}c^{2})(\mu_{B}^{e}H)}}~.
\end{equation} 
Obviously, there is a distinct different behavior of both solutions with
respect to the magnetic field, which is expected to be
 more pronounced at larger magnetic field strengths.

\vspace{3mm}
\noindent
For a numerical illustration, let us consider the case for
  a field of 1 Tesla. Utilizing the Bohr magneton of
  an electron (in Gauss units) 
  \footnote{$erg=0.624 \cdot 10^{12}$ eV, $10^{4}$ Gauss = 1 Tesla.} 
\begin{eqnarray}
     \mu_{B}^{e} = \frac{e\hbar}{2m_{e}c} =
      0.927 ~10^{-20} ~erg ~Gauss^{-1} = 
      5.788 ~10^{-9} ~eV ~Gauss^{-1} = \nonumber\\
     = 5.788 ~10^{-15}~MeV ~ Gauss^{-1} =   5.788 ~10^{-11} ~MeV ~T^{-1}
\end{eqnarray}          
and eq.(3), the following value
 for the nonrelativistic transverse energy of the
 first exited state is obtianed
\begin{equation}
      E_{T,\lambda=-\frac{1}{2}}^{nonrel}(n=1)_{H=1T} =
                     2\mu_{B}^{e}H= \nonumber\\
        2\cdot 5.788 ~10^{-9} ~eV ~Gauss^{-1} ~10^{4} ~Gauss  
        \approx  1.158  ~10^{-4} ~eV .
\end{equation}
  If this number for
  $E_{T,\lambda=-\frac{1}{2}}^{nonrel}(n=1)_{H=1T}$ is introduced
  into (15 ), the following value
  for the relativistic transverse energy of an electron
  being in the same magnetic field with identical
  quantum numbers $n$ and  $\lambda$ is found
\begin{equation}
{E_{T,\lambda=-\frac{1}{2}}(n=1)_{H=1T}  \approx  10.87 ~eV~. }
\end{equation}
     Comparing the numbers in (17) and (18), one notices that
   for the magnetic
   field  strength  of 1 Tesla the  relativistic
   Dirac equation provides for the first radial excitation
   a transverse energy \textit{five orders of magnitude}
   higher than the nonrelativistic Schr\"odinger  equation. 
   Technically, this mismatch can be understood from
   formula (15) because a) the square root of $H$
    enlarges $E_{T,\lambda}$
   by about two orders of magnitude and b) the second
    multiplication  factor,  $ \sqrt { 2(m_{e}c^{2})}$
   (with $m_{e}c^{2} = 0.511 ~MeV=~0.511\cdot 10^{6} ~eV$),
   increases the energy level by additional three
   orders of magnitude. 
   The substantial difference derived for a magnetic
   field of 1 Tesla, being expected to grow
   with increasing magnetic field strength, was to
   our knowledge never discussed in the literature so far.

\vspace{3mm}
\noindent   
   In this connection it is of interest to recall that
  the situation with the physical interpretation of the
  exact solutions of the Dirac equation is not so 
  definite in cases of strong electric fields.
  For example, it is well known that  the solution
  for the energy levels of the Dirac equation using the
  Coulomb potential,   $V(r) = - \frac{Ze^{2}}{r}$, 
  where the charge factor Z 
  defines  the strength of the electric
  field,  can be written as  
\begin{eqnarray}
   E^{Coul.rel}_{n,j} = 
    mc^{2}(1 + \frac{(\alpha{Z})^{2} }
     { (n - (j +1/2)+\sqrt{(j+1/2)^2-(\alpha Z)^2}~)^2 } )^{-1/2}~,
\end{eqnarray} 
   with  
   $\alpha=\frac{e^{2}}{\hbar c}=\frac{1}{137}$ and
    $ n=j + 1/2 + k = 1,2,...$ as the main quantum
    number, and  $j = 1/2$  for $l=0$ or
    $j= l \pm 1/2$ if $l \not= 0 $.
    Eq.(19) has however a restricted range
    of physical validity. 
  For instance,  for the smallest value $j=1/2$ the
  expression under the 
  square root in the
  dominator  becomes negative and leads to 
  unphysical solutions if $Z > Z_{cr} $, with $Z_{cr}=137$
  as the critical
  value for the charge factor. 
   
\vspace{3mm}
\noindent
  In contrast to the relativistic expression (19),
  the analogous formula for the bound state energy levels 
  within the Schr\"odinger equation
\begin{eqnarray}
   E^{Coul.nonrel}_{n,j} = - \frac{R\hbar{Z^{2}}}{n^{2}}~,
\end{eqnarray} 
  with the Rydberg constant $R=me^{4}/2\hbar^{3}$
  and $n = 1,2,...$, is valid for all Z values.
  To resolve experimentally the discrepancy between the relativistic
 solution (19) and the nonrelativistic expression (20),
 in particular for large values of Z,
 is challenging since pointlike charges with
 $Z \ge 137$ do not exist in nature \footnote{See, e.g.
            \cite{Akhi} -\cite{Bagr_etal}
            for more discussions of this problem 
            in connection of heavy ion cases.}.
  
\vspace{3mm}
\noindent
  Unlike the example of the Coulomb potential,
  the solution of the Dirac equation (6)
  for transverse energy levels of electrons 
  within in a magnetic field is valid
  for any strength of the field H and, thereby,
  the experimental
  verification of the predictions (3), respectively, 
  (6) is possible even at very high field strengths
     \footnote{We have not yet found any results 
               of such a study in the literature.}.
     
\vspace{3mm}
\noindent
    Such a task may be performed by passing an electron beam
    through a static homogeneous field of about 0.1-10 Tesla.
    Electrons of up to 100 MeV (which are considerd
    to be relativistic due to the smallness of their mass)
    may occupy some quasi-stable quantized energy levels
    $E_{T,\lambda}(n)$. For large life times of these levels,
    see \cite{Zioutas} and \cite{Ashkin},
    transitions of excited electrons to the ground state (10)
    are limited during passing through the H-field and
    registration of emitted photons should be performed
    sufficiently downstream of the magnet.
    
\vspace{3mm}
\noindent
Also, "stimulation" of beam electrons by absorption of laser light
inside a magnet leads to
"exited"  states of electrons which may be followed by
    emission of $\gamma$-rays within or behind the magnet
    depending on the lifetime of the excited states.
    Details  of such an experiment should however be considered
    as soon as its realization becomes appropriate.

%
\section{Summary}
%
  Numerical values for the transverse energy of
  electrons which propagate in  a static homogeneous
  magnetic field were calculated using the
  relativistic (Dirac) and the  nonrelativistic (Schr\"odinger)
  equations. Employing a magnetic field of 1 Tesla and
  non-spinflip transitions from orbits with $n=1$ to $n=0$, 
  as an example,  a difference of five orders of magnitude 
  between the relativistic
  and nonrelativistic concepts 
  for the electrons' transverse energy
  was evaluated.
  In other words, electrons traversing
  a magnetic field of 1 Tesla radiate photons
  being  about $10^5$ times more energetic in the 
  relativistic than in the
  nonrelativistic case, for non-spinflip transitions
  from $n=1$ to $n=0$ orbits.
%
   
\vspace{3mm}
\noindent
  Finally, we believe that experimental verification
  of the predictions from either the relativistic or
  the nonrelativistic equation  on
  quasi-atomic quantized  energy levels of electrons
  traversing a strong static magnetic field  would
  be very desirable.
  Such measurements can be performed by studying, 
  for example, Compton scattering of laser light with
  electrons when both beams move parallel along
  the magnetic field lines. Registration
  of radiated  photons, caused by
  electron transitions  from higher 
  to lower orbits, should resolve the difference
  between the relativistic  and nonrelativistic predictions
  and should also provide a good test of how
  to add the interaction terms to the Dirac equation.

\subsection*{Acknowledgment}

\noindent
We would like to thank Desmond Barber for helpful discussions
and reading of the manuscript.


\end{document}